\documentclass[12pt]{article}
\usepackage{graphicx}
\usepackage{epstopdf}
\begin{document}

\def\a{\alpha}
\def\b{\beta}
\def\c{\varepsilon}
\def\d{\delta}
\def\e{\epsilon}
\def\f{\phi}
\def\g{\gamma}
\def\h{\theta}
\def\k{\kappa}
\def\l{\lambda}
\def\m{\mu}
\def\n{\nu}
\def\p{\psi}
\def\q{\partial}
\def\r{\rho}
\def\s{\sigma}
\def\t{\tau}
\def\u{\upsilon}
\def\v{\varphi}
\def\w{\omega}
\def\x{\xi}
\def\y{\eta}
\def\z{\zeta}
\def\D{\Delta}
\def\G{\Gamma}
\def\H{\Theta}
\def\L{\Lambda}
\def\F{\Phi}
\def\P{\Psi}
\def\S{\Sigma}

\def\o{\over}
\def\beq{\begin{eqnarray}}
\def\eeq{\end{eqnarray}}
\newcommand{\gsim}{ \mathop{}_{\textstyle \sim}^{\textstyle >} }
\newcommand{\lsim}{ \mathop{}_{\textstyle \sim}^{\textstyle <} }
\newcommand{\vev}[1]{ \left\langle {#1} \right\rangle }
\newcommand{\bra}[1]{ \langle {#1} | }
\newcommand{\ket}[1]{ | {#1} \rangle }
\newcommand{\EV}{ {\rm eV} }
\newcommand{\KEV}{ {\rm keV} }
\newcommand{\MEV}{ {\rm MeV} }
\newcommand{\GEV}{ {\rm GeV} }
\newcommand{\TEV}{ {\rm TeV} }
\def\diag{\mathop{\rm diag}\nolimits}
\def\Spin{\mathop{\rm Spin}}
\def\SO{\mathop{\rm SO}}
\def\O{\mathop{\rm O}}
\def\SU{\mathop{\rm SU}}
\def\U{\mathop{\rm U}}
\def\Sp{\mathop{\rm Sp}}
\def\SL{\mathop{\rm SL}}
\def\tr{\mathop{\rm tr}}

\def\IJMP{Int.~J.~Mod.~Phys. }
\def\MPL{Mod.~Phys.~Lett. }
\def\NP{Nucl.~Phys. }
\def\PL{Phys.~Lett. }
\def\PR{Phys.~Rev. }
\def\PRL{Phys.~Rev.~Lett. }
\def\PTP{Prog.~Theor.~Phys. }
\def\ZP{Z.~Phys. }

\newcommand{\bea}{\begin{eqnarray}}   
\newcommand{\eea}{\end{eqnarray}}
\newcommand{\bear}{\begin{array}}  
\newcommand {\eear}{\end{array}}
\newcommand{\la}{\left\langle}  
\newcommand{\ra}{\right\rangle}
\newcommand{\non}{\nonumber}  
\newcommand{\ds}{\displaystyle}
\newcommand{\red}{\textcolor{red}}
\def\ubl{U(1)$_{\rm B-L}$}
\def\REF#1{(\ref{#1})}
\def\lrf#1#2{ \left(\frac{#1}{#2}\right)}
\def\lrfp#1#2#3{ \left(\frac{#1}{#2} \right)^{#3}}
\def\OG#1{ {\cal O}(#1){\rm\,GeV}}

\newcommand{\ah}{A_H}

\baselineskip 0.7cm

\begin{titlepage}

\begin{flushright}
IPMU 09-0126
\end{flushright}

\vskip 1.35cm
\begin{center}
{\large \bf
Gamma-ray Constraints on Hadronic and Leptonic Activities of Decaying Dark Matter
}
\vskip 1.2cm
Chuan-Ren Chen$^{1}$, Sourav K. Mandal$^{1,2}$, Fuminobu Takahashi$^{1}$
\vskip 0.4cm

{\it 
$^1$ Institute for the Physics and Mathematics of the Universe, 
University of Tokyo,\\ Chiba 277-8568, Japan\\
$^2$ Department of Physics, University of California,\\ Berkeley, CA 94720, USA\\
}

\vskip 1.5cm

\abstract{ While the excess in cosmic-ray electrons and positrons
  reported by PAMELA and Fermi may be explained by dark matter
  decaying primarily into charged leptons, this does not necessarily
  mean that dark matter should not have any hadronic decay modes.  In
  order to quantify the allowed hadronic activities, we derive
  constraints on the decay rates of dark matter into $WW$, $ZZ$,
  $hh$, $q\bar{q}$ and $gg$ using the Fermi and HESS gamma-ray data.
  We also derive gamma-ray constraints on the leptonic $e^+e^-$,
  $\mu^+\mu^-$ and $\tau^+\tau^-$ final states.  We find that dark
  matter must decay primarily into $\mu^+\mu^-$ or $\tau^+\tau^-$ in
  order to simultaneously explain the reported excess and meet all
  gamma-ray constraints.  }
\end{center}
\end{titlepage}

\setcounter{page}{2}

\section{Introduction}
\label{sec:1}
The existence of dark matter (DM) has been firmly established by
numerous observations, though the nature of DM mostly remains
unknown. In particular, it is not known whether DM is absolutely
stable or not.  If DM is unstable, it will eventually decay into
lighter particles which may be observed as an excess in the cosmic-ray
spectrum.

If DM is related to new physics which appears at the weak scale, it is
natural to expect that the DM mass is in the range ${\cal
  O}(100)\,{\rm GeV}$--${\cal O}(10)$\,TeV.  However, the longevity of DM
whose mass is of the weak scale is a puzzle and calls for some
explanation.  The (quasi)stability may be the result of a discrete
symmetry or extremely weak interactions. For instance, in a
supersymmetric (SUSY) theory, the lightest SUSY particle (LSP) is
stable and therefore a candidate for DM if R-parity is an exact
symmetry. However, R-parity violation may be a common phenomenon in
the string landscape~\cite{Kuriyama:2008pv}, in which case the LSP DM
is unstable and eventually decays into Standard Model (SM) particles.
On the other hand, if DM is in a hidden sector which has extremely
suppressed interactions with the SM sector, the only way to probe DM
may be to look in the cosmic rays for signatures of its decay
products.

Recently, much attention has been given to the electron/positron
excess reported by PAMELA~\cite{Adriani:2008zr}, ATIC~\cite{ATIC},
PPB-BETS~\cite{Torii:2008xu} and Fermi~\cite{Abdo:2009zk}.  The excess
clearly suggests that we need to modify our current understanding of
the production/acceleration/propagation of cosmic-ray electrons and
positrons. Of the many explanations proposed so far for this excess,
DM decay or annihilation remains an exciting possibility.  In order to
account for the excess by DM decay/annihilation, DM should mainly
produce leptons with suppressed hadronic branching ratios, otherwise
the anti-protons produced would likely exceed the observed
flux~\cite{Adriani:2008zq}. Further model-independent analysis also
revealed that, if one requires that the PAMELA/Fermi excess be
explained by DM annihilation, the gamma-rays accompanying the lepton
production exceeds the observed flux unless a significantly less steep
DM profile is assumed~\cite{Bertone:2008xr,meade:fermi}.  Thus, the
leptophilic decaying DM scenario has recently gained
momentum~\cite{Chen:2008yi,Chen:2008dh,Yin:2008bs,
  Ishiwata:2008cv,Shirai:2009kh,Ibarra:2009bm,Ibe:2009en}.

Leptophilic decaying DM models can be broadly divided into two
categories.  One is such that DM first decays into additional light
particles, which subsequently decay into muons or electrons, while the
decays into hadrons are forbidden by
kinematics~\cite{Cholis:2008vb}. The other is such that the DM
particle couples mainly to leptons due to symmetry~\cite{Chen:2008dh}
or geometric setup~\cite{Okada:2009bz}. While the hadronic activities
are absent in the former case, it is model-dependent to what extent
the DM is leptophilic in the latter case.  One example is the hidden
gauge boson decaying into the SM particles through a mixing with a
U(1)$_{B-L}$ gauge boson~\cite{Chen:2008yi}; the DM is certainly
leptophilic in the sense that it mainly decays into leptons, but a
certain amount of quarks are also produced.

The purpose of this paper is to study the current constraints on the
hadronic and leptonic decay of DM. The anti-proton flux is known to
provide a tight constraint on the hadronic activities, but there are
large uncertainties in the propagation~\cite{Hisano:2005ec}. The other
constraint comes from gamma-ray observation. In contrast to charged
cosmic-ray particles, gamma-rays travel undeflected and there is no
uncertainty in the propagation; the main uncertainty is the dark
matter profile.  Furthermore, the Fermi satellite has been measuring
gamma-rays with both unprecedented precision and statistics, and we
can expect a significant improvement over EGRET
data~\cite{Sreekumar:1997un,Strong:2004ry}.  In this paper we will
derive constraints on the partial decay rates of DM into $WW$,
$ZZ$, $hh$, $q\bar{q}$ and $gg$ as well as $e^+e^-$, $\mu^+\mu^-$ and
$\tau^+\tau^-$ using the Fermi~\cite{Fermi:ML, Fermi:iso} and
HESS~\cite{HESS:GC} data.  The bounds obtained in this paper are not
only generic but also can be used to know what branching ratios are
allowed in decaying DM models which account for the PAMELA/Fermi
cosmic-ray electron/positron excess.

\section{Analysis}
\label{sec:2}
There are several contributions to the gamma-ray spectrum when DM
decays into SM particles. Photons from fragmentation are generated by
the decay of mesons, especially $\pi^0$, and final-state radiation
(FSR) is always produced when the DM decays into charged
particles. The electrons and positrons produced by the DM decay will
lose energy by emitting synchrotron radiation in the galactic magnetic
field and through inverse Compton (IC) scattering off ambient photons
(star light, dust re-emission, and CMB).  In this section we summarize
the calculation of these contributions and how we derive constraints.

\subsection{Local contributions}
First, we use {\tt PYTHIA 6.4.21}~\cite{Sjostrand:2006za} to simulate
the fragmentation of the various final states at a range of DM masses,
and for charged final states we use the expression in 
Ref.~\cite{FSR} for the photon multiplicity from
final-state radiation.
Since these contributions are local to the site of DM decay, and
because gamma-rays travel undeflected, the differential flux from our
galactic DM halo is given by flux conservation: 
\beq
\left(\frac{d\Phi_\gamma}{dEd\Omega}\right)^{\rm (gal.)}_{\rm local}
=\frac{1}{4\pi}(r_\odot\rho_\odot)\frac{1}
{m_{\rm DM} \tau}
\left(\frac{dN^{\rm (frag.)}_\gamma}{dE}+\frac{dN^{\rm (FSR)}_\gamma}{dE}\right)
J(\Delta\Omega)
\eeq
where $r_\odot=8.5$\,kpc is the solar distance from the galactic
center~\cite{Catena:2009mf}, $\rho_\odot=0.3\,{\rm GeV}\,{\rm
  cm}^{-3}$ is the density of the DM halo at this distance, $\tau$
is the DM lifetime, and 
\beq
J(\Delta\Omega)=\frac{1}{\Delta\Omega}\int_{\Delta\Omega} d\Omega
\int_{\rm l.o.s.}\frac{ds}{r_\odot}\left(\frac{\rho_{\rm DM}(\vec x)}{\rho_\odot}\right)
\eeq
is the line-of-sight integral, in which
$\Delta\Omega$ is the region of sky observed by a given
experiment and $\rho_{\rm DM}(\vec x)$ is the halo profile.  
In this analysis we use the NFW halo
profile~\cite{Navarro:1996gj}.

There is also an isotropic, diffuse extragalactic contribution on
cosmological scales from DM residing in our past light cone.  We find
that this contribution to the differential flux is given by
\beq
\left(\frac{d\Phi_\gamma}{dEd\Omega}\right)^{\rm (ex.)}_{\rm local}
=\frac{c}{4\pi}
\frac{\Omega_{\rm DM}\rho_c}{H_0\Omega_{\rm M}^{1/2}}\frac{1}{m_{\rm DM} \tau}
\int_1^\infty dy 
\frac{y^{-3/2}}{\sqrt{1+\Omega_\Lambda/\Omega_{\rm M}y^{-3}}}
\left(\frac{dN^{\rm (frag.)}_\gamma}{d(Ey)}+\frac{dN^{\rm (FSR)}_\gamma}{d(Ey)}\right)
\eeq
where $y\equiv 1+z$ and the cosmological 
parameters are given by Ref.~\cite{WMAP5};
the density of radiation is taken to be
negligible.  This expression duplicates the result of
Ref. \cite{Ibarra:exgal}.

\subsection{Contributions from propagating electrons and positrons}
Electrons and positrons as final states of DM decay, as well as those
from the fragmentation of other final states, will lose energy via
synchrotron radiation and IC scattering off ambient photons.  Here we
describe the calculation of these effects and the resulting
contribution to the gamma-ray flux; our analysis parallels that of
Ref. \cite{moroi:prop}.

First, let us consider the galactic contribution.  The diffusion of
$e^\pm$ is governed by the equation
\beq
K(E)\nabla^2f_e(E,\vec x)+\frac{\partial}{\partial E}
\left[b(E, \vec x)f_e(E,\vec x)\right]+Q(E,\vec x)=0
\eeq
where $K(E)$ is the diffusion coefficient, $f_e(E, \vec x)$ is the
$e^\pm$ phase space distribution, $b(E, \vec x)=b_{\rm syn}(E, \vec
x)+b_{\rm IC}(E, \vec x)$ is the energy loss rate, and $Q(E, \vec x$)
is the $e^\pm$ source term.  
For this analysis we adopt
the MED propagation model~\cite{Delahaye:2007fr} to set $K(E)$ and the
geometric boundary of diffusion.  Under this model the propagation
length for $e^\pm$ with $E\gsim 100$\,GeV is ${\cal O}(0.1)$\,kpc
before losing its most of the energy.  In the limit that this length
is small compared to the distance traveled, $f_e(E,\vec x)$ is
well-approximated by
\beq
f_e(E,\vec x)=\frac{1}{b(E,\vec x)}\frac{\rho_{\rm DM}(\vec x)}{m_{\rm DM} \tau}
\int_E^\infty dE'\, \frac{dN_e}{dE'}\;.
\eeq
The energy loss rate due to synchrotron radiation is given by
\beq
b_{\rm syn}(E)=\frac{4}{3}\sigma_T \left(\frac{E}{m_e}\right)^2
\left(\frac{B^2}{2}\right)
\eeq
where $\sigma_T$ is the Thomson cross-section and $B\approx3\mu{\rm
  G}$ is taken as the strength of the 
  galactic magnetic field.\footnote{The value of the 
magnetic field strength may be different close to the
galactic center.}  The
energy loss rate due to IC scattering is given by \beq b_{\rm IC}(E,
\vec x)=\int dE_\gamma dE_{\gamma_{\rm BG}} (E_\gamma-E_{\gamma_{\rm
    BG}})\frac{d\sigma_{\rm IC}}{dE_\gamma} f_{\gamma_{\rm
    BG}}(E_{\gamma_{\rm BG}}, \vec x) \eeq where $d\sigma_{\rm
  IC}/dE_\gamma$ is the IC differential cross-section as given in
Ref. \cite{icscs}, and $f_{\gamma_{\rm BG}}(E_{\gamma_{\rm BG}}, \vec
x)$ is the sum of the CMB radiation field and the galactic radiation
field (star light and dust re-emission).  In our calculation we use the
interstellar radiation field (ISRF) furnished by the GALPROP
collaboration~\cite{galprop:light}.

Because the magnetic field is so weak, synchrotron emissions would
only be in the radio and is thus only relevant in this calculation as
an energy-loss mechanism.  On the other hand, IC processes for $e^\pm$
of $E\sim{\cal O}(100)$\,GeV would produce gamma-rays of $E\sim{\cal
  O}(1)$\,GeV which is in the range of the Fermi observations.
Therefore, the differential gamma-ray flux is given by
\beq
\left(\frac{d\Phi_\gamma}{dEd\Omega}\right)^{\rm (gal.)}_{\rm prop.}
=
\frac{1}{4\pi}\frac{1}{\Delta\Omega}\int_{\Delta\Omega}d\Omega
\int_{\rm l.o.s.}ds 
\int dE_e dE_{\gamma_{\rm BG}}\,\frac{d\sigma_{\rm IC}}{dE}
f_{\gamma_{\rm BG}}(E_{\gamma_{\rm BG}},\vec x)f_e(E_e,\vec x)\;.
\eeq

Now let us turn to the extragalactic contribution.  On cosmological
distances, there is no interstellar radiation field, only the CMB, and
negligibly small magnetic fields.  Thus, the only energy-loss
mechanism is IC scattering of CMB photons.  Moreover, assuming the
universe is indeed isotropic and homogeneous, the distributions of CMB
photons and DM are spatially invariant.  Then the diffusion equation
becomes \beq \frac{\partial}{\partial E}\left[b(t, E)f_e(t,
  E)\right]+Q(t, E) +H\,E\;\frac{\partial f_e(t, E)}{\partial E} =
\frac{\partial f_e(t, E)}{\partial t} \eeq where $H$ is the Hubble
parameter.  Since the CMB photon energy is so low the $e^\pm$ are
non-relativistic in the center-of-momentum frame, the energy loss rate
due to IC scattering is given by the Thomson limit
\beq
b_{\rm cosm}(y,E)=\frac{4}{3}\left(\frac{E}{m_e}\right)^2
\sigma_T\,\left(\rho_{\rm CMB}\,y^4\right)
\eeq
where $\rho_{\rm CMB}\simeq 0.26\,{\rm eV}\,{\rm cm}^{-3}$ is the
present-day CMB energy density and $y\equiv1+z$ as before.  The
timescale of energy-loss $E/b_{\rm cosm}\lsim{\cal O}(10^{14})\,$sec
is much less than the Hubble time, so the term ${\cal O}(H)$ in the
diffusion equation can be ignored.  This gives for the $e^\pm$
spectrum
\beq
f_e(y, E)=\frac{1}{b_{\rm cosm}(y, E)}\;\frac{1}{m_{\rm DM} \tau}\;
\left(\rho_{\rm DM}\,y^3\right)
\int_E^\infty dE'\; \frac{dN_e}{dE'}\;.
\eeq
Finally, we find for the differential flux
\beq
\left(\frac{d\Phi_\gamma}{dEd\Omega}\right)^{\rm (ex.)}_{\rm prop.}
=\frac{c}{4\pi}
\frac{1}{H_0\Omega_{\rm M}^{1/2}}\int_1^\infty dy 
\frac{y^{-9/2}}{\sqrt{1+\Omega_\Lambda/\Omega_{\rm M}y^{-3}}} A(y, E)
\eeq
where
\beq
A(y, E)\equiv\int dE_e\,dE_{\gamma_{\rm BG}}\;\frac{d\sigma_{\rm IC}}{d(Ey)}\;
f_{\gamma_{\rm BG}}(y,E_{\gamma_{\rm BG}})\;f_e(y,E_e)
\eeq
and $f_{\gamma_{\rm BG}}(y,E_{\gamma_{\rm BG}})$ is the CMB spectrum at
redshift $z=y-1$.
Comparable expressions for annihilating dark matter may be found
in Refs.~\cite{belikov:cmbic,profumo:cmbic}.

\subsection{Derivation of constraints from data sets}
We derive constraints using three data sets: observation of gamma-rays
from the galactic center (within $0.1^\circ$) 
by the HESS telescope~\cite{HESS:GC}, observation of the
galactic mid-latitude ($10^\circ\leq|b|\leq20^\circ$) diffuse
gamma-ray flux by the Fermi LAT~\cite{Fermi:ML}, and preliminary data
for the isotropic diffuse flux ($|b|\geq 10^\circ)$ also by the Fermi
LAT~\cite{Fermi:iso}.

We do not use the HESS observation of the
galactic center ridge~\cite{HESS:GR} because it requires a subtraction
of nearby flux levels.\footnote{We are grateful to 
J.~F.~Beacom~\cite{Beacom:GR} for pointing this out.}  The result of this
procedure is highly profile dependent: we find that 
for the NFW profile this procedure weakens the constraint by a factor 5, and 
for the less-cuspy isothermal profile the constraint would be weakened by a factor
200.  Before the procedure our result is comparable to that of
Ref.~\cite{meade:fermi}.

For various DM masses and lifetimes, we calculate and sum the various
contributions to the gamma-ray flux for each of the energy bins of
each data set, with two exceptions due to limited computational power:
\begin{itemize}
\item We use only the two highest bins of the Fermi
  mid-latitude data (energy $\sim{\cal O}(10)$\,GeV) when calculating
  the constraints for weak boson and colored channels. 
  Since the spectra from DM decay are harder than the observed
  spectra, any excess will
  be dominated by the highest energy bins anyway.
\item We include galactic ICS for comparison to the HESS galactic center 
data only for the leptonic final states, because
  they copiously produce hard $e^\pm$.  Also, since we
  cannot know at this time how much galactic contribution (namely, galactic ICS
  and galactic halo contributions) is present in the
  isotropic flux reported by Fermi, we give this constraint
  both with and without galactic contribution added in.  For the Fermi mid-latitude
  data we include galactic ICS for all channels.
\end{itemize}

\begin{figure}[t!]
  \begin{center}
    \includegraphics[width=0.4\linewidth]{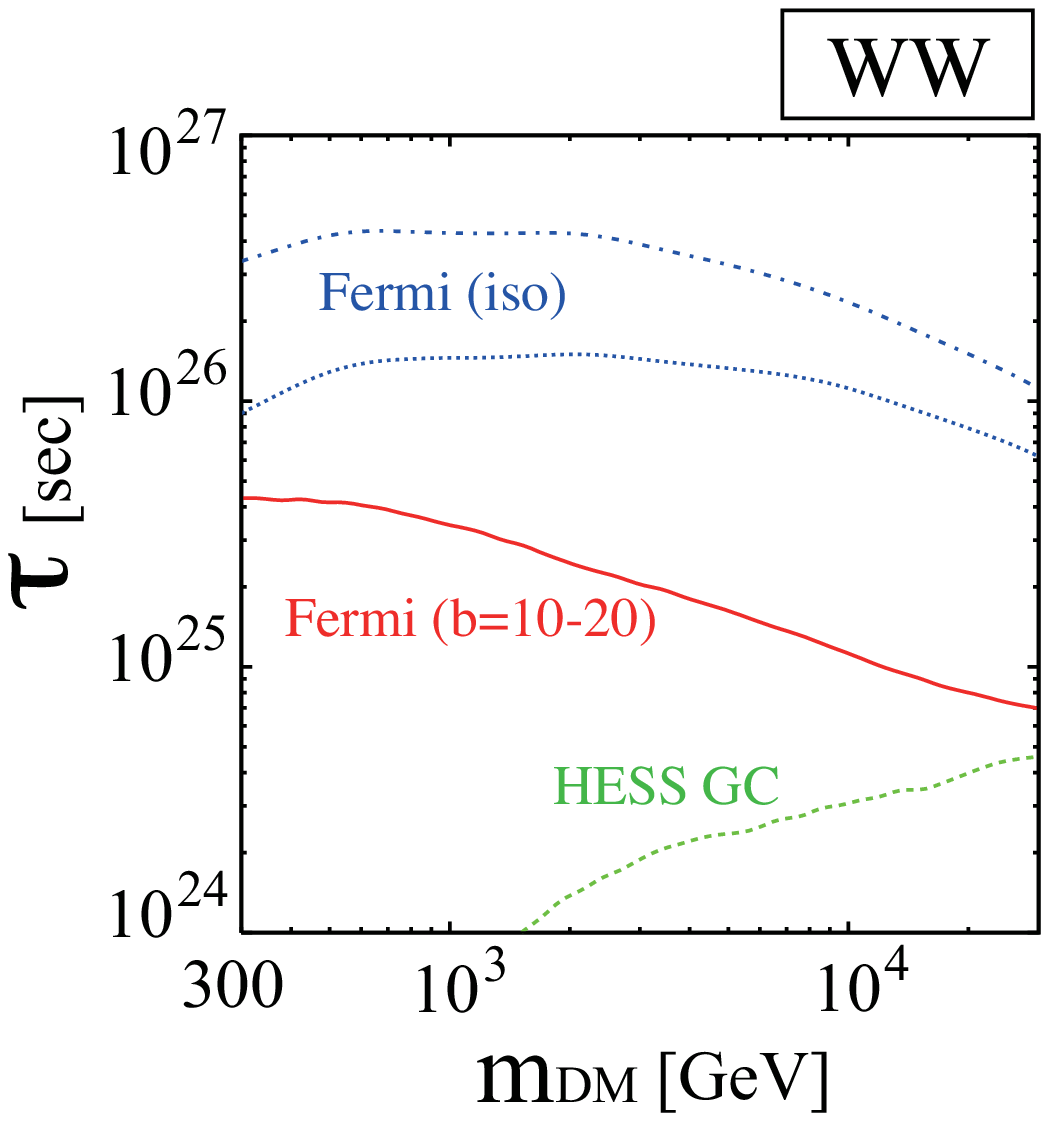}
    \includegraphics[width=0.4\linewidth]{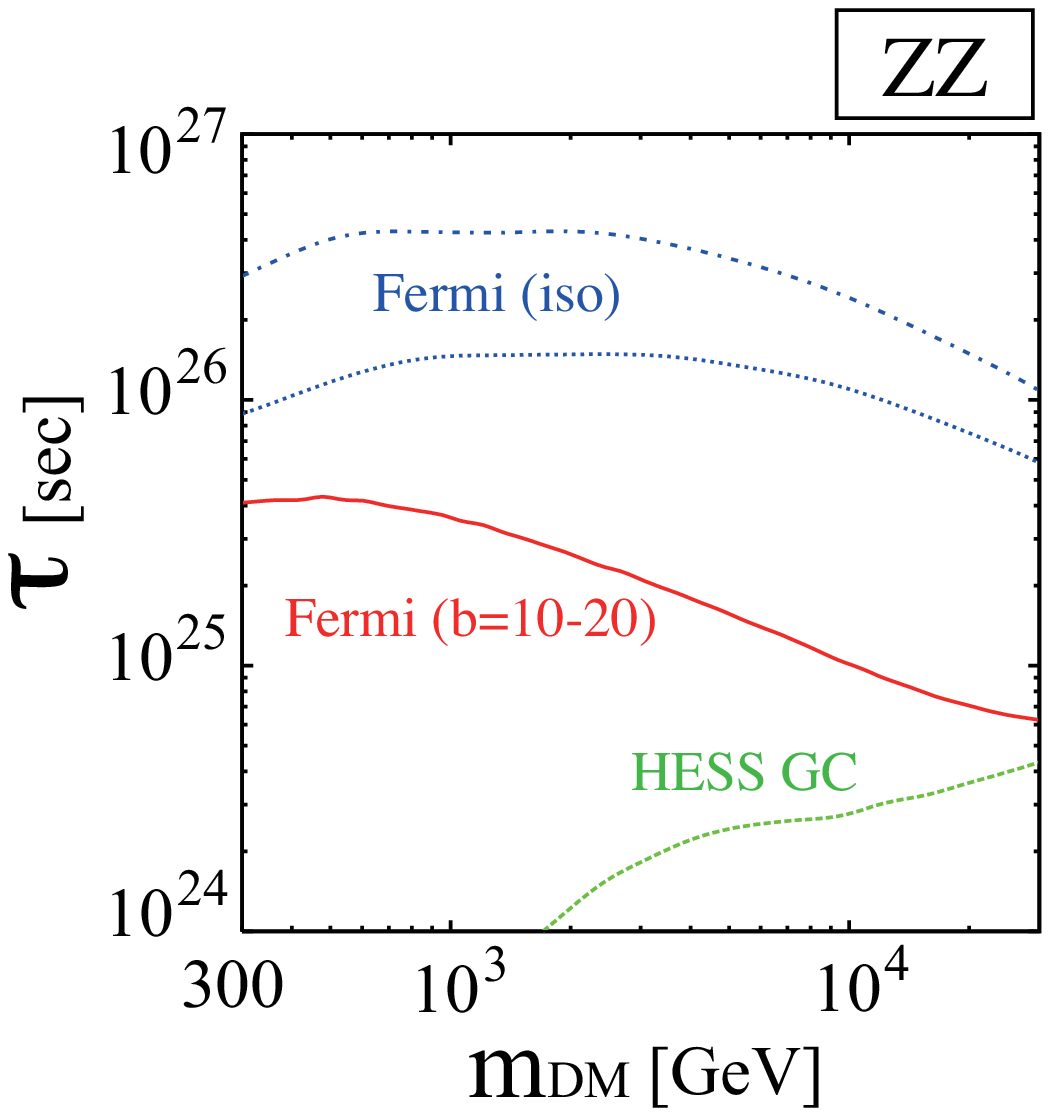}
    \includegraphics[width=0.4\linewidth]{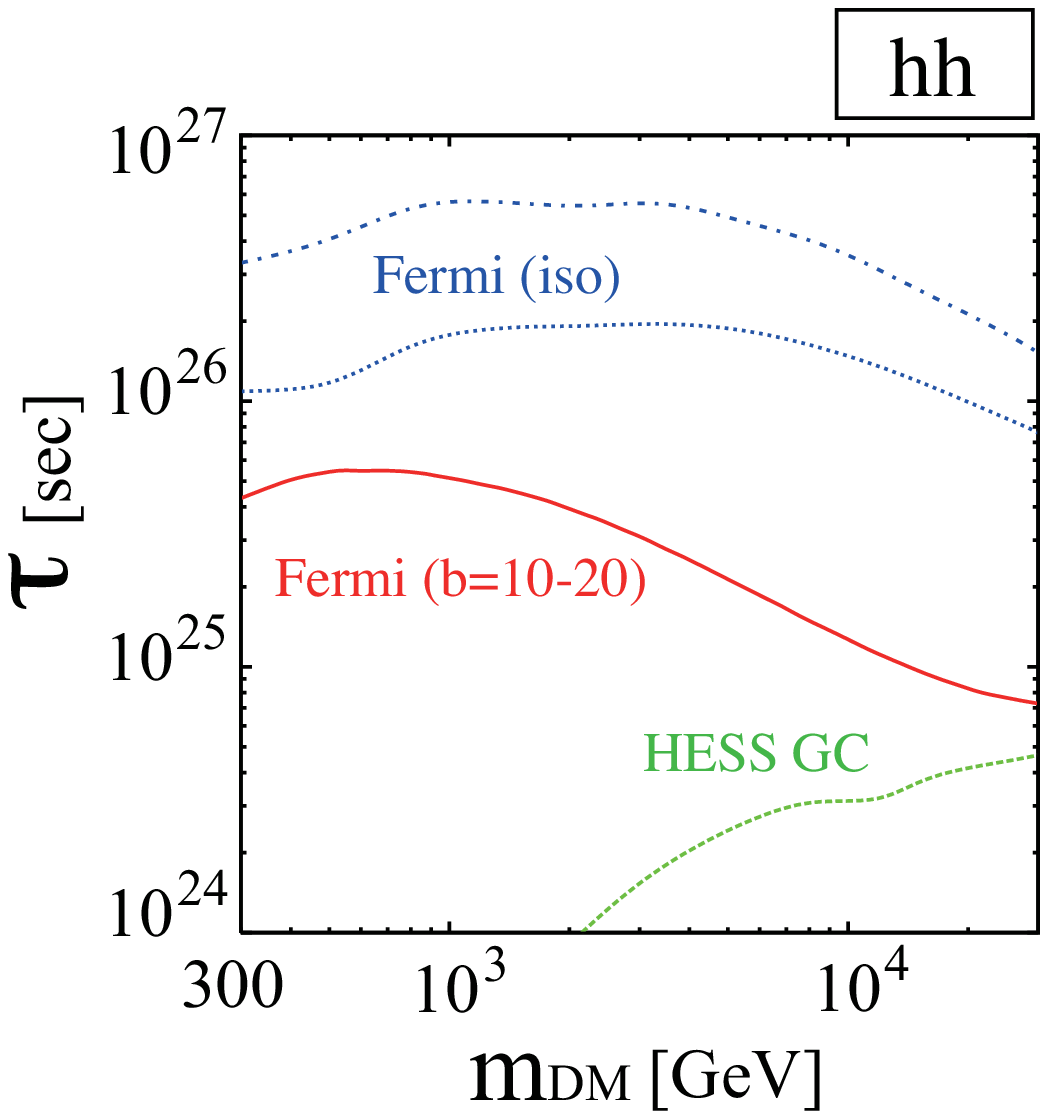}
    \caption{3$\sigma$ constraints for $WW$, $ZZ$ and $hh$ final states.
    The lower dotted blue
line is the constraint from Fermi isotropic diffuse flux without galactic contribution,
and the upper dotted-dashed blue line is the the constraint with galactic contribution.
    }
    \label{fig:weak}
  \end{center}
\end{figure}

\begin{figure}[t]
  \begin{center}
	\includegraphics[width=0.4\linewidth]{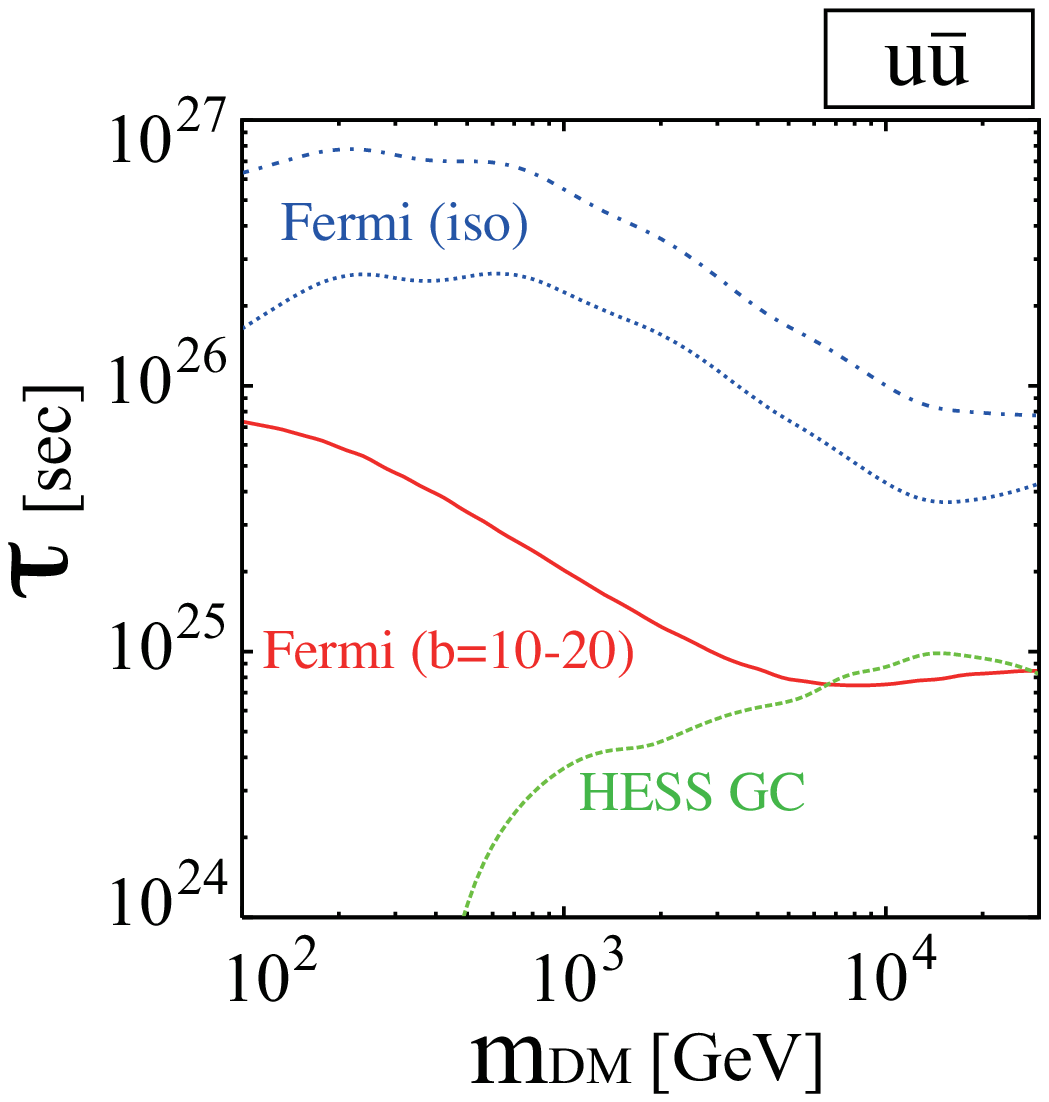}
    \includegraphics[width=0.4\linewidth]{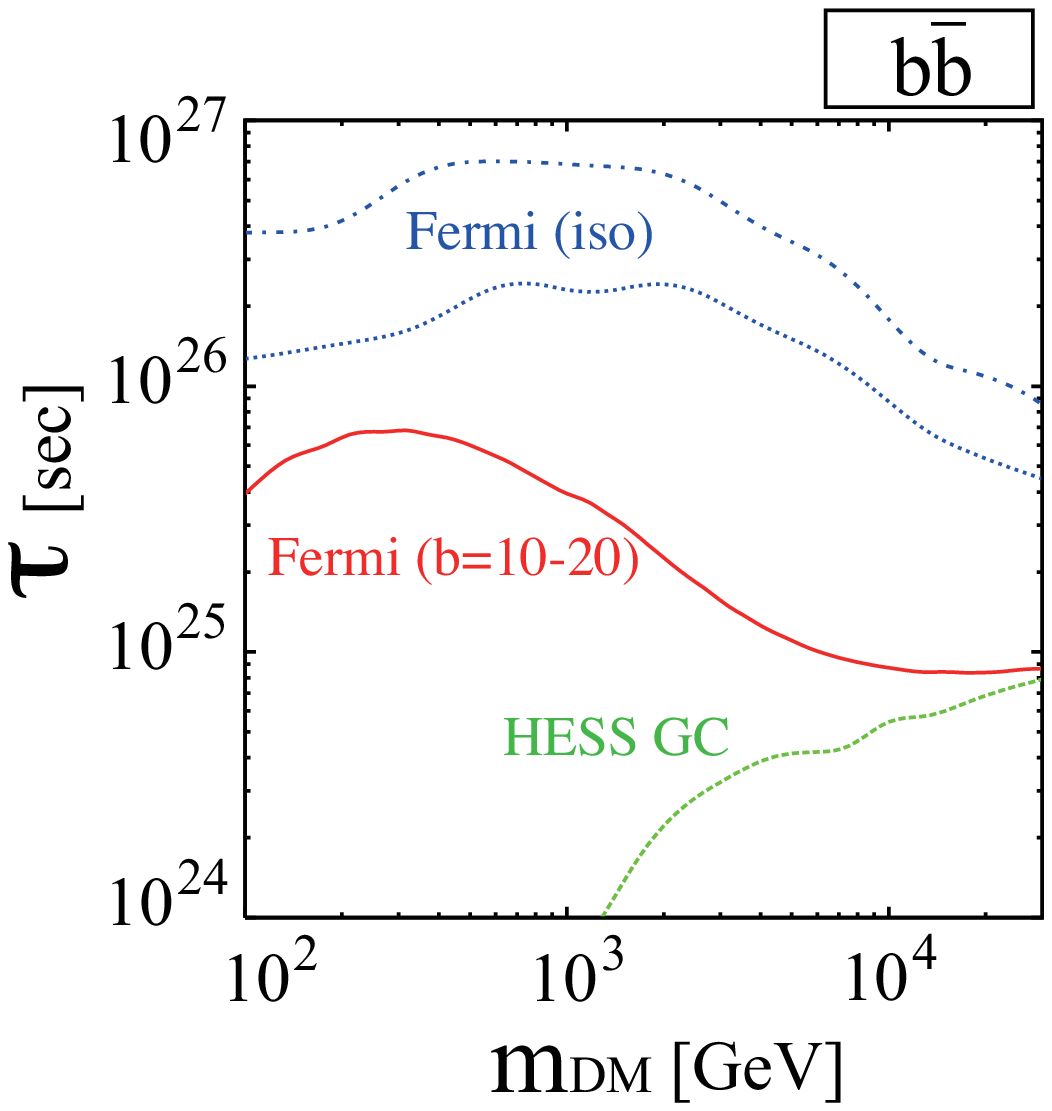}
    \includegraphics[width=0.4\linewidth]{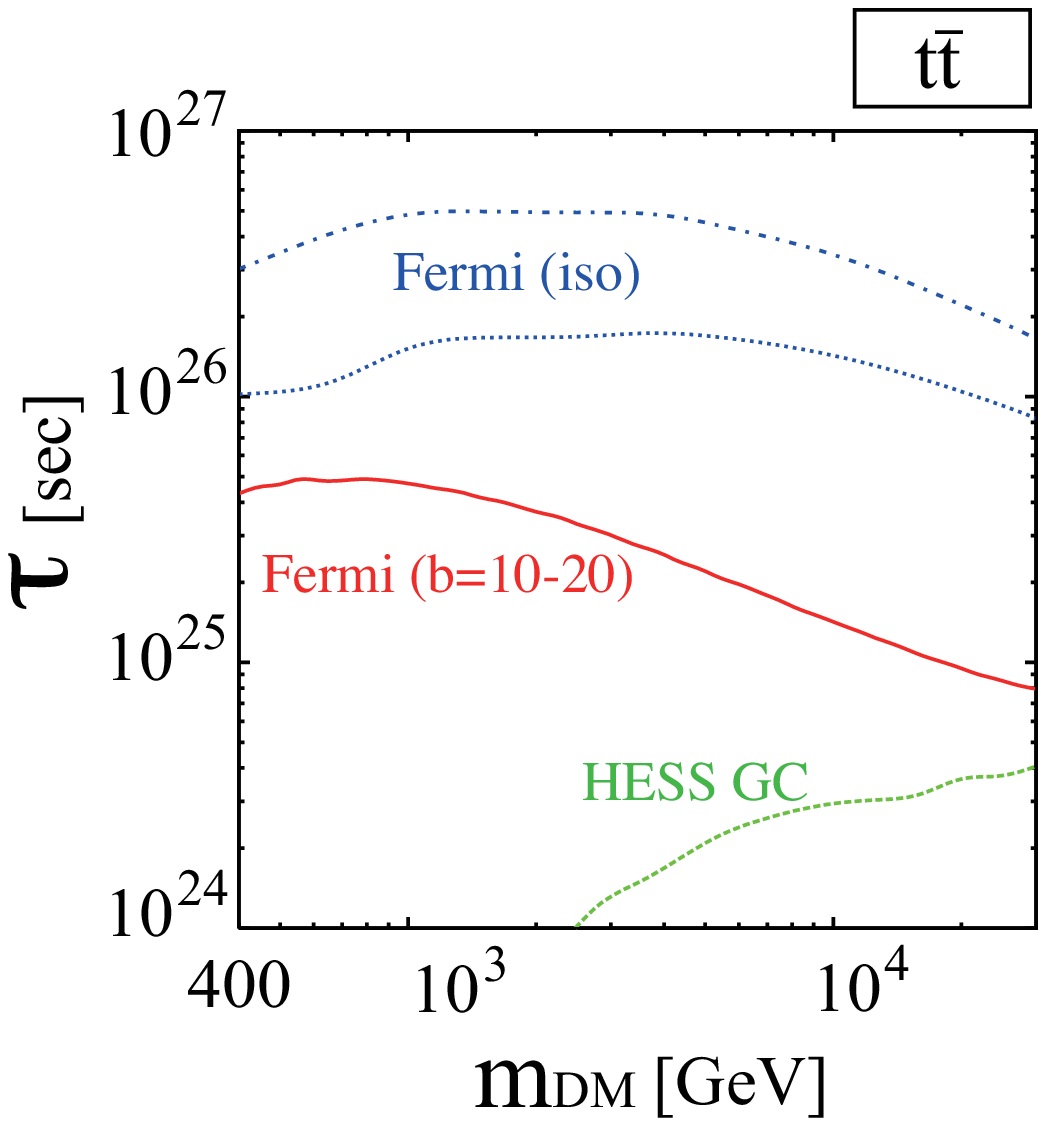}
    \includegraphics[width=0.4\linewidth]{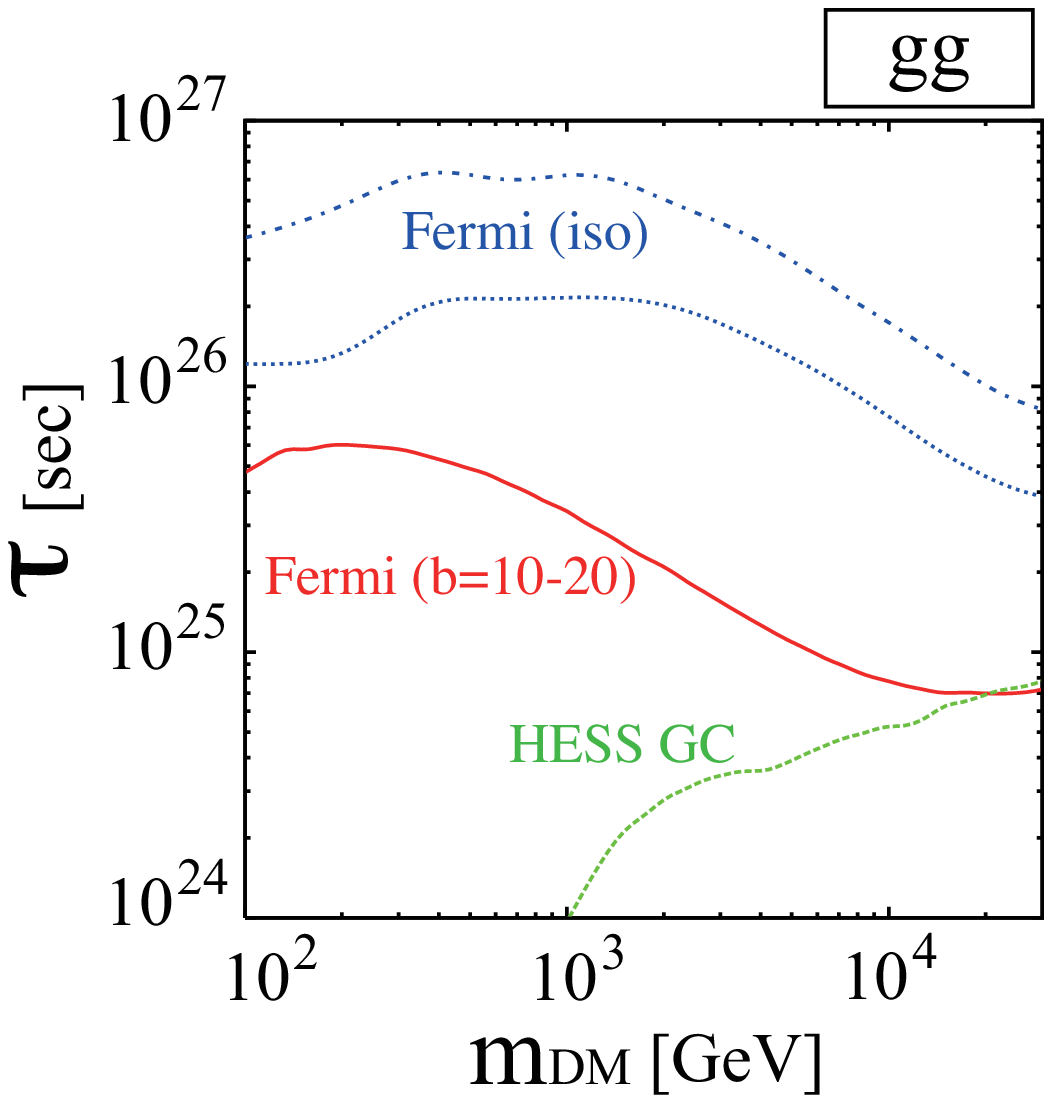}
    \caption{3$\sigma$ constraints for colored final states.}
    \label{fig:colored}
  \end{center}
\end{figure}

\begin{figure}[t!]
  \begin{center}
    \includegraphics[width=0.4\linewidth]{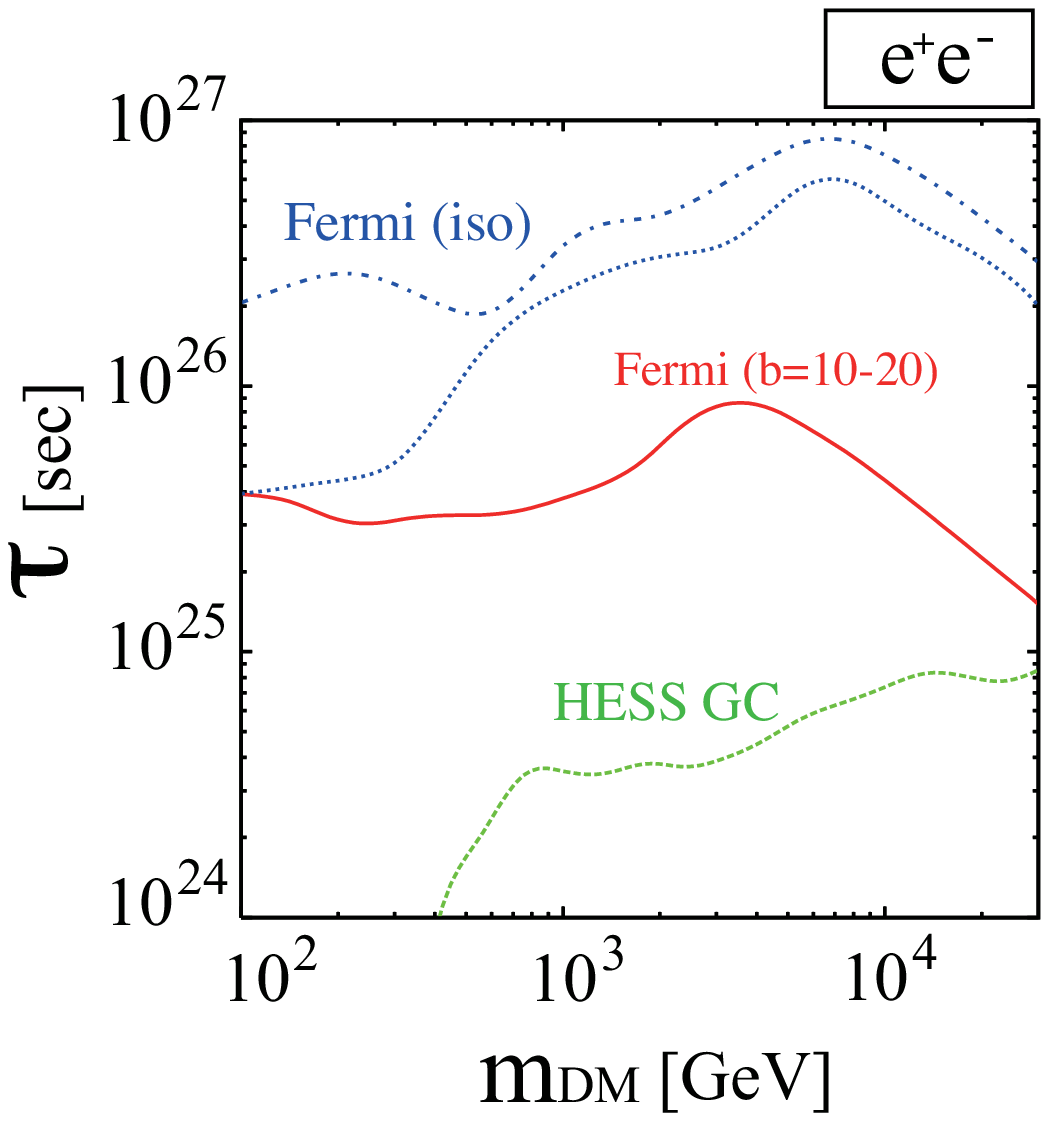}
    \includegraphics[width=0.4\linewidth]{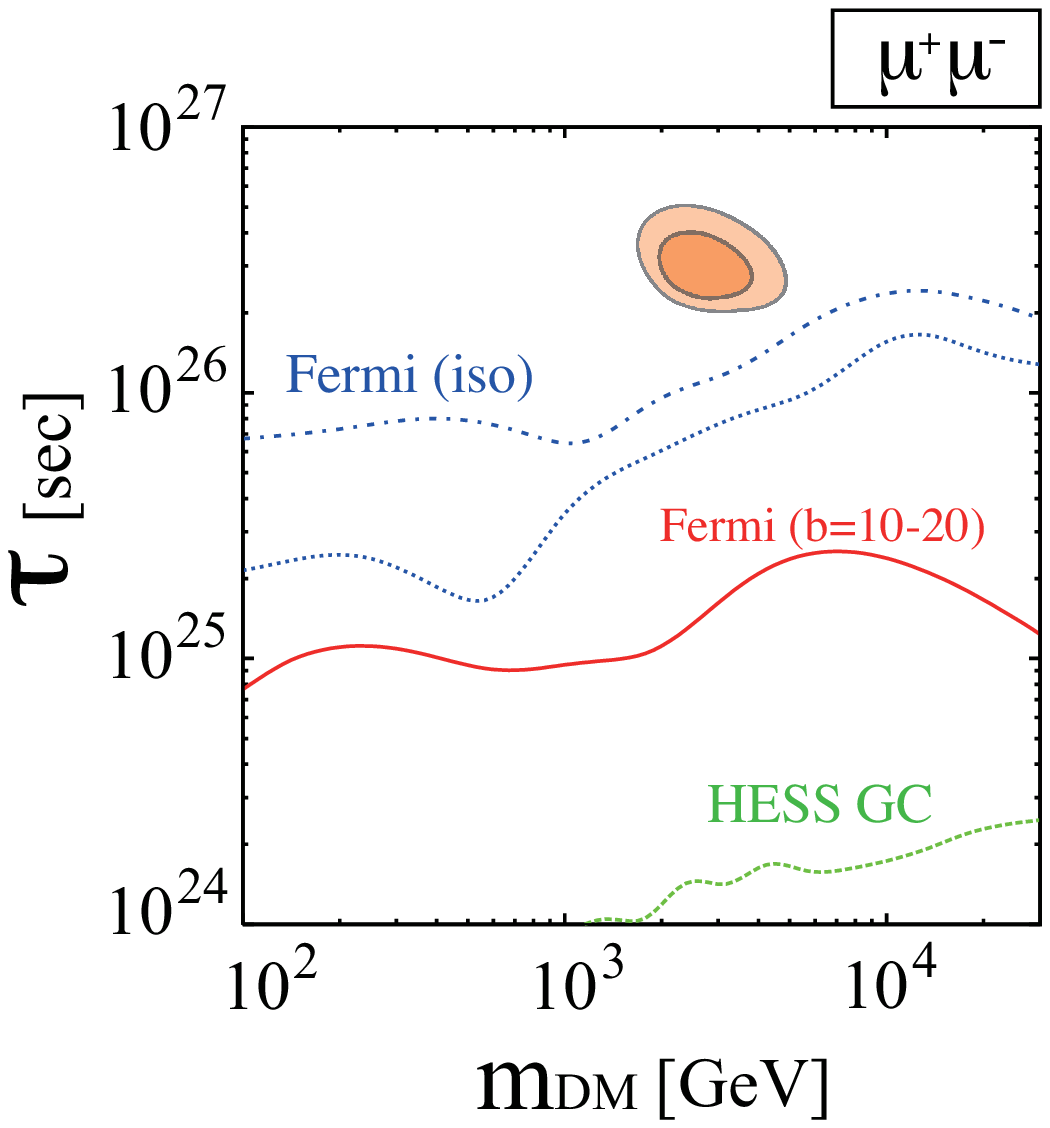}
    \includegraphics[width=0.4\linewidth]{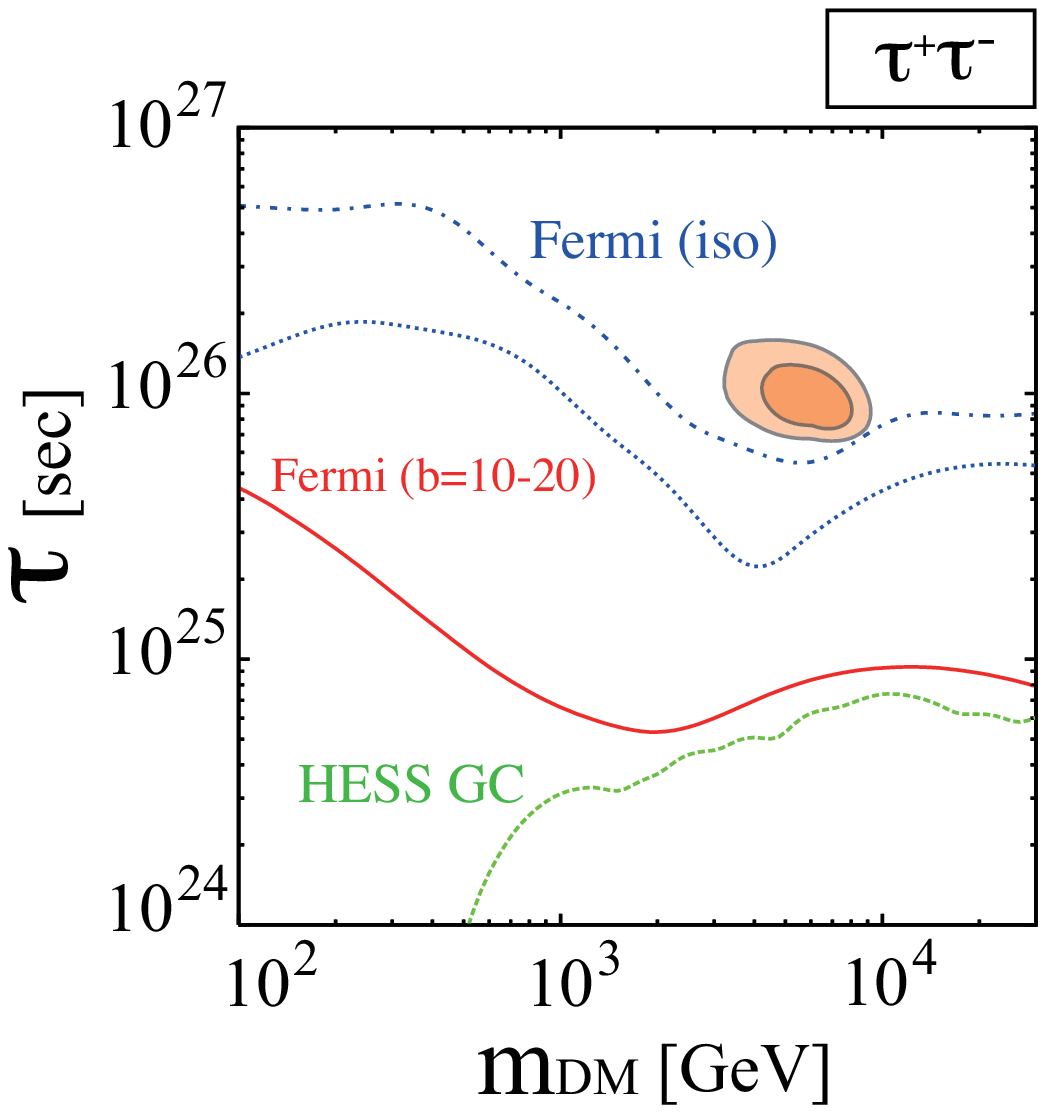}	
    \caption{
    3$\sigma$ constraints for leptonic final states.  
The orange ellipse represents the region favored by the 
PAMELA/Fermi excess~\cite{meade:fermi}.
}
    \label{fig:leptonic}
  \end{center}
\end{figure}

Then, we compute how many standard deviations the calculated flux
exceeds the data for these bins and take the largest of these, as in
Ref. \cite{meade:fermi}.  Again because the spectra from DM decay
are harder than the observed spectra, 
this statistic is little different than reduced-$\chi^2$, but
without the aliasing errors due to the number of effective degrees of
freedom changing near a contour.  The 3$\sigma$ contours are shown on
Figures \ref{fig:weak}, \ref{fig:colored}, and \ref{fig:leptonic} for
weak and Higgs boson, colored and leptonic channels, respectively.
The lower dotted blue line is the constraint
from Fermi isotropic diffuse flux without galactic contribution, and the upper
dotted-dashed blue line is the the constraint with galactic contribution.

\subsection{Uncertainties}
Finally, let us consider the uncertainties in these calculations:
\begin{itemize}
\item {\it DM halo profile:} We use the NFW profile for the galactic
  dark matter halo.  If we were to use instead the isothermal
  profile, the calculated flux of fragmentation and FSR photons from
  the galactic center would be reduced by a factor 4, weakening
  the constraint from the HESS observation.  By contrast, because the
  Fermi observations cover a much larger portion of the sky looking
  away from the galactic center, switching to the isothermal profile
  would hardly change these constraints.
\item {\it Local DM density:} We use the conservative, standard value 
$\rho_\odot=0.3\,{\rm GeV}\,{\rm cm}^{-3}$ for the DM halo density at the 
radius of our solar system, though a wide range is accepted \cite{PDG2008}.  A recent
 analysis gives the value $\rho_\odot=0.39\,{\rm GeV}\,{\rm cm}^{-3}$ 
within 10\%~\cite{Catena:2009mf}.
\item {\it ICS calculation:} There are uncertainties associated with parameters
such as the 
galactic magnetic field and interstellar radiation field, as well as the 
choice of the diffusion model and the approximation of dropping the
diffusion effect in calculating the 
$e^\pm$ phase space distribution.  Ref.~\cite{ibarra:ics} found that the same
calculation with a slightly different setup matched the numerical
results of GALPROP, and Ref.~\cite{meade:fermi} estimates that 
the no-diffusion approximation can change photon flux by~a~factor~of~2.  
Consequently, we estimate that the uncertainty in our
calculation of ICS is ${\cal O}(1)$.
\end{itemize}

\section{Discussion}
For a whole range of DM masses, it is the Fermi isotropic diffuse flux
that places most stringent constraints on the lifetime, followed by the
Fermi galactic mid-latitude data and then by the HESS galactic center
data.  We can also see that for the hadronic channels shown in 
Figs.~\ref{fig:weak} and \ref{fig:colored},
as well as for the $\tau^+\tau^-$ channel shown in Fig.~\ref{fig:leptonic},
 the constraints derived from the Fermi mid-latitude 
data becomes weaker at heavier
DM masses.  This feature can be understood as follows. The gamma-ray
energy spectrum from decaying DM is dominated by fragmentation and
final-state radiation at energies near the threshold $\sim m_{\rm
  DM}/2$, while the up-scattered photons through IC process dominate
at low energies. Since the Fermi mid-latitude 
data is available up to ${\cal O}(10)$\,GeV,
it depends on the DM mass which contribution becomes important. The
up-scattered photons tend to be dominant at $E \simeq {\cal O}(10)$\,GeV for
a DM mass above ${\cal O}(1)$\,TeV. That is why the constraints
become flatter above ${\cal O}(1)$\,TeV.  The constraints from the
Fermi isotropic data, which also goes up to ${\cal O}(10)$\,GeV,
 show a similar flattening for $u\bar u$, $b\bar b$, $gg$ and 
 $\tau^+\tau^-$ final states, which fragment into hard mesons.  It is interesting to note
that the IC scattering is important for the hadronic and $\tau^+\tau^-$
modes if the DM mass is heavier than several TeV.

From Fig.~\ref{fig:leptonic} one can see that the decay into $e^+e^-$
and $\tau^+\tau^-$ are more tightly constrained than that into
$\mu^+\mu^-$. This is because the direct production of energetic
electron/positron enhances the gamma-ray flux through final-state
radiation and IC scattering,
and because the decay of $\tau$ is accompanied by the fragmentation
photons.  The DM lifetime should be ${\cal O}(10^{26})$ sec in order to
explain the PAMELA/Fermi excess in cosmic-ray electrons/positrons
(shown by the orange ellipses in the figure).  Therefore, in order to
simultaneously satisfy the gamma-ray constraints and as well as
account for the PAMELA and Fermi excess, DM must decay primarily into
either muons or taus.

Suppose that the DM mainly decays into $\mu^+\mu^-$ at a lifetime of
${\cal O}(10^{26})$ sec.  The allowed hadronic branching ratio is about
${\cal O}(10)$\% from the figure. Of course the precise value depends on DM
models. In next section we consider a DM model in which a hidden gauge
boson decays into quarks and leptons through a mixing with an
U(1)$_{B-L}$ gauge boson.  The allowed hadronic branching ratio in
this DM model is about $30\%$ at a reference point, $m_{\rm DM} =
2$\,TeV and $\tau = 1.5 \times 10^{26}$ sec.

We may, however, compare these generic gamma-ray constraints with the
generic constraints from neutrinos.  Bounds on leptonic final states
from upward-going muons due to neutrinos interacting with the
Earth~\cite{superk:muon} are 10--100 times weaker than those
presented here, though the direct neutrino bounds from
IceCube+DeepCore will surpass present-day gamma-ray constraints after
a few years of running~\cite{deepcore}.

In the above analysis we required that the the DM contribution should
not exceed the Fermi/HESS data points at more than $3 \sigma$.  If a
fraction $f$ of the observed flux is due to the astrophysical
gamma-ray sources, the constraint on the DM lifetime will be improved
by about $1/f$.  For instance, Ref.~\cite{Inoue:2008pk} proposed a
model of blazars and estimate its contribution to the diffuse
isotropic gamma-ray flux. Although there is uncertainty in the blazar
model, their estimate agrees well with the preliminary Fermi data. As
the astrophysical understanding of the origin of the observed
gamma-rays flux is improved, the constraint on the DM property gets
stronger.

\section{Constraints on a dark matter model}
\label{sec:3}
Using the constraints derived in the previous section, we should be
able to check whether a specific DM model is allowed by the current
gamma-ray observation.  As an example, we take up a model which was
proposed in Ref.~\cite{Chen:2008md} to account for the PAMELA and
Fermi excess.

First, let us briefly review the model (see the original reference for
details).  The lifetime needed to account for the excess is of
${\cal O}(10^{26})$ seconds, and such a longevity calls for some
explanation. To this end we introduce an extra dimension, which is
assumed to be compactified on $S^1/Z_2$ with two distinct boundaries.
Suppose that a hidden U(1)$_{\rm H}$ gauge field is confined on one
boundary and the SM particles on the other. In such a set-up, direct
interactions between the two sectors are suppressed by a factor of
$\exp({- M_* L})$, where $M_*$ is the five-dimensional Planck scale
and $L$ denotes the size of the extra dimension. For e.g. $M_* L \sim
10^2$, the direct couplings are so suppressed that the hidden gauge
boson will be practically stable in a cosmological time
scale~\cite{Chen:2008md}. Assuming that the hidden U(1)$_{\rm H}$
gauge symmetry is spontaneously broken, the hidden gauge boson, $A_H$,
can be therefore a candidate for DM.

Let us introduce an U(1)$_{\rm B-L}$ in the bulk.  Through a kinematic
mixing between the U(1)$_{\rm H}$ and U(1)$_{\rm B-L}$, which is
generically present, the $A_H$ will then decay into the SM quarks and
leptons at a rate determined by their $B-L$ quantum number.  After
integrating out the heavy U(1)$_{\rm B-L}$ gauge boson\footnote{We
  expect that the U(1)$_{\rm B-L}$ gauge symmetry is spontaneously
  broken around the grand unification theory (GUT) scale of about
  $10^{15}$ GeV, since the seesaw mechanism~\cite{seesaw} for neutrino
  mass generation suggests the right-handed neutrinos of a mass about
  $10^{15}$ GeV.}, the effective couplings between the hidden gauge
boson $\ah$ and the SM fermion $\psi_i$ can be extracted from the
U(1)$_{\rm B-L}$ gauge interactions
\beq
{\cal L_{\rm int}} \;=\;  
-\lambda\, q_i \frac{m^2}{M^2}  A_H^{\mu} \, \bar{\psi}_i \gamma_\mu \psi_i,
\eeq
where $\lambda$ is a coefficient of the kinetic mixing, $q_i$ denotes
the $B-L$ charge of the fermion $\psi_i$, and $m$ and $M$ are the
masses of the hidden gauge boson $A_H$ and the U(1)$_{\rm B-L}$ gauge
boson, respectively. The lifetime of $\ah$ is therefore given as
\beq
\tau \;\simeq\; 1\times 10^{25} {\rm \,sec} \left(\sum_i N_i q_i^2\right)^{-1}
\lrfp{\lambda}{0.01}{-2}
 \lrfp{m}{2{\rm\, TeV}}{-5} \lrfp{M}{10^{15}{\rm GeV}}{4},
\eeq
where $N_i$ is the color factor ($3$ for quarks and $1$ for leptons),
the sum is taken over the SM fermions, and we have neglected the
fermion masses.  The introduction of the U(1)$_{\rm B-L}$ has two
merits. One is that, for a natural choice of the $B-L$ breaking scale,
the lifetime of DM falls in a desired range of ${\cal O}(10^{26})$
seconds. The other is that the DM decay mode is leptophilic and the
branching ratios simply reflect the $B-L$ charge assignment, which
makes the model very predictive; the branching ratio into a quark pair
is given by $2/39$, while that into a charged lepton pair is $2/13$.

\begin{figure}[t]
  \begin{center}
    \includegraphics[width=0.6\linewidth]{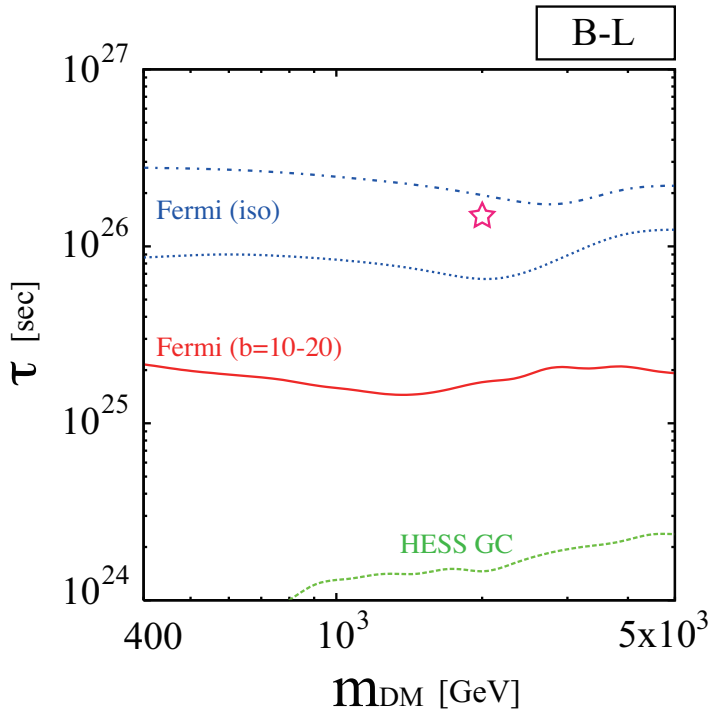}
    \caption{3$\sigma$ constraints for U(1)$_{\rm B-L}$ model with a
      star representing the reference point, $m_{\rm DM} =
      2{\rm\,TeV}$ and $\tau = 1.5 \times 10^{26}$ sec, which accounts
      for both PAMELA and Fermi excess.}
    \label{fig:bminusl}
  \end{center}
\end{figure}

In a similar way as we did in the previous section, we have derived
constraints on the lifetime of $\ah$ from the gamma-ray data.  See
Fig.~\ref{fig:bminusl}.  The Fermi galactic mid-latitude data plays an
important role in constraining the model at low side of DM masses,
while the HESS galactic center data takes over for the high side.  For
example, the lifetime of the $\ah$ of mass $500$ GeV ($2$ TeV) should
be longer than about $2 \times 10^{25}$ ($2.7 \times 10^{25}$)
seconds. We take the reference point $m_{\rm DM} = 2$\,TeV and the
lifetime $\tau = 1.5 \times 10^{26}$\,sec shown as a star in
Fig.~\ref{fig:bminusl} which can explain both the Fermi and PAMELA
excess. The antiproton flux was calculated for the reference point and
we found that it is consistent with the current
PAMELA~\cite{Adriani:2008zq} and other observational data on the
cosmic-ray antiproton.  On the other hand, as one can see from
Fig.~\ref{fig:bminusl}, the reference point is marginally excluded by the
bound obtained from the Fermi isotropic
diffuse data. Since the gamma-ray flux calculation is more robust over
the antiproton flux\footnote{ The flux of the antiproton produced by
  DM annihilation/decay is very sensitive to the diffusion model and
  it varies by about two orders of magnitude for different diffusion
  models.\cite{Hisano:2005ec} }, we conclude that the hidden gauge
boson DM model, which can account for both the PAMELA and Fermi excess
while satisfying the antiproton flux constraint, is now marginally excluded by
the Fermi isotropic diffuse gamma-ray data.
 
\section{Conclusions}
\label{sec:5}
In this paper we have derived constraints on the partial decay rates
of DM into $WW$, $ZZ$, $hh$ and $q\bar{q}$ as well as $e^+e^-$,
$\mu^+\mu^-$ and $\tau^+\tau^-$ using the gamma-ray data observed by
the Fermi LAT and HESS. One of the merits of using gamma-ray is that
the predicted flux does not depend on the diffusion model in the
Galaxy, in contrast to charged cosmic-rays. The constraints derived in
this paper provide implications for DM model-building attempting to
account for the PAMELA/Fermi excess.  According to our results, the
allowed hadronic branching ratio is of ${\cal O}(10)$ \%.  We have applied
the result to a DM model based on the hidden gauge boson decaying
through a mixing with the U(1)$_{B-L}$, and found that the model is
now marginally excluded by the Fermi gamma-ray observation. The allowed
hadronic branching ratio is about $30\%$ at the reference point shown 
as a star in Fig.~\ref{fig:bminusl}.

Gamma-ray observations have the power to constrain the properties of
DM.  Thanks especially to the Fermi LAT data, which provides greater
accuracy and statistics over the experiments in the past, the
gamma-ray constraint has become as tight as or even tighter than
current neutrino and antiproton constraints.  Further observation by
the Fermi satellite will hopefully shed light on the dark sector of
our Universe.

\section*{Acknowledgments}
CRC and FT are grateful to T.~Yanagida for stimulating discussion.
FT thanks S.~Shirai for a useful communication concerning the numerical calculation.
This work was supported by World Premier International Center
Initiative (WPI Program), MEXT, Japan.  FT was supported by JSPS
Grant-in-Aid for Young Scientists (B) (21740160) and the Grant-in-Aid for 
Scientific Research on Innovative Areas (No. 21111006).

\end{document}